\documentclass[twocolumn,showpacs,preprintnumbers,amsmath,amssymb]{revtex4}
\usepackage{graphicx}
\usepackage{dcolumn}
\usepackage{bm}

\begin{document}
\preprint{APS/123-QED}

\title{Model ingredients and peak mass production in heavy-ion collisions}

\author{Sukhjit Kaur$^{1}$ }
\author{Aman D. Sood$^{2,}$}%
\email{amandsood@gmail.com} \affiliation{$^{1}$House No. 465,
Sector-1B, Nasrali, Mandi Gobindgarh-147301, Punjab, India}
\affiliation{$^{2}$SUBATECH, Laboratoire de Physique Subatomique et des Technologies Associ\'{e}es, Universit\'{e} de Nantes - IN2P3/CNRS - EMN \\
4 rue Alfred Kastler, F-44072 Nantes, France.}


\date{\today}

\begin{abstract}
We simulate the central reactions of $^{20}$Ne+$^{20}$Ne,
$^{40}$Ar+$^{45}$Sc, $^{58}$Ni+$^{58}$Ni, $^{86}$Kr+$^{93}$Nb,
$^{129}$Xe+$^{118}$Sn, $^{86}$Kr+$^{197}$Au, and
$^{197}$Au+$^{197}$Au at different incident energies for different
equations of state, different binary cross sections and different
widths of Gaussians. A rise-and-fall behavior of the multiplicity
of intermediate mass fragments (IMFs) is observed. The system size
dependence of peak center-of-mass energy E$_{c.m.} ^{max}$ and
peak IMF multiplicity $\langle$N$_{IMF}\rangle^{max}$ is also
studied, where it is observed that E$_{c.m.}^{max}$ follows a
linear behavior and $\langle$N$_{IMF}\rangle^{max}$ shows a
power-law dependence. A comparison between two clusterization
methods, the minimum spanning tree and the minimum spanning tree
method with binding energy check (MSTB), is also made. We find
that the MSTB method reduces the $\langle$N$_{IMF}\rangle^{max}$,
especially in heavy systems. The power-law dependence is also
observed for fragments of different sizes at E$_{c.m.} ^{max}$ and
the power-law parameter $\tau$ is found to be close to unity in
all cases except A$^{max}$.
\end{abstract}


\maketitle

\section{Introduction}
 At high excitation energies, the colliding nuclei may break into several small
and intermediate size fragments and a large number of nucleons are
also emitted \cite{rkpuri,sood09,jsingh00}. The emission of
intermediate mass fragments (IMFs) in nuclear collisions was
studied for more than a decade during which several experimental
groups have carried out a complete study of fragment formation
with 4$\pi$ detectors
\cite{blaich93,tsang93,sisan01,peas94,desouza91,stone97,ogil91}.
These studies revealed that the fragments formed in heavy-ion
collisions depend crucially on the bombarding energy and impact
parameter of the reaction
\cite{rkpuri,sood09,jsingh00,blaich93,tsang93}. Therefore, these
experimental studies of fragmentation offer a unique opportunity
to explore the mechanism behind the formation of the fragments.
Moreover, one can also pin down the role of dynamics in fragment
formation and its time scale.
\par
Recently, there has been increasing interest in the effects of
reaction dynamics on the production of IMFs and light charged
particles (LCPs, $\emph{Z}$ = 1 or 2). Sisan $\textit{et al.}$
\cite{sisan01} studied the emission of IMFs from central
collisions of nearly symmetric systems using a 4$\pi$-array set
up, where they found that the multiplicity of IMFs shows a rise
and fall with increase in the beam energy. They observed that
E$_{c.m.}^{max}$ (the energy at which the maximum production of
IMFs occurs) increases linearly with the system mass, whereas a
power-law ($\propto$ A$^{\tau}$) dependence was reported for peak
multiplicity of IMFs with power factor $\tau$ = 0.7. Peaslee
$\textit{et al.}$ \cite{peas94}, however, studied the asymmetric
system $^{84}$Kr+$^{197}$Au in the incident energy range from 35
to 400 MeV/nucleon and obtained an energy dependence of
multifragmentation. Their findings revealed that fragment
production increases up to 100 MeV/nucleon and then decreases with
increase in incident energy. De Souza $\textit{et al.}$
\cite{desouza91} studied the central collisions of
$^{36}$Ar+$^{197}$Au from 35 to 120 MeV/nucleon and observed that
IMF multiplicity shows a steady increase with increase in the
incident energy. The IMF multiplicity decreases, however, when one
moves from central to peripheral collisions. However, Tsang
$\textit{et al.}$ \cite{tsang93}, in their investigation of
$^{197}$Au+$^{197}$Au collisions at E/A = 100, 250, and 400 MeV,
found the occurrence of peak multiplicity at lower energies for
central collisions whereas it is shifted to higher energies for
peripheral collisions. Stone $\textit{et al.}$ \cite{stone97} used
a nearly symmetric system of $^{86}$Kr+$^{93}$Nb from 35 to 95
MeV/nucleon to obtain IMF multiplicity distribution as a function
of beam energy by selecting central events. Ogilvie $\textit{et
al.}$ \cite{ogil91} also studied the multifragment decays of Au
projectiles after collisions with C, Al, and Cu targets at the
bombarding energy of 600 MeV/nucleon using the ALADIN forward
spectrometer at GSI, Darmstadt, with the beam accelerated by
Schwerionensynchrotron (SIS). They found that, with increasing
violence of collision, the mean multiplicity of IMFs originating
from the projectile first increases to a maximum and then
decreases again. As mentioned earlier, Sisan $\textit{et al.}$
\cite{sisan01} reported that the peak multiplicity of IMFs as well
as peak center-of-mass energy scale with the size of the system.
In a recent communication, Vermani and Puri \cite{puri09}
succeeded partially in explaining the above-mentioned behavior by
using the quantum molecular dynamics (QMD) approach. Here we plan
to extend the above study by incorporating various model
ingredients such as equation of state, nucleon-nucleon (nn) cross
section, and Gaussian width. The role of different clusterization
algorithms is also explored. We attempt to find out whether these
ingredients have sizable effects.
\par
\section{The Formalism}
\subsection{Quantum molecular dynamics model}
\par
We describe the time evolution of a heavy-ion reaction within the
framework of a QMD model \cite{rkpuri,sood09,jsingh00,aich91},
which is based on a molecular dynamics picture. Here each nucleon
is represented by a coherent state of the form
\begin{equation}
\phi_{\alpha}(x_1,t)=\left({\frac {2}{L \pi}}\right)^{\frac
{3}{4}} e^{-(x_1-x_{\alpha }(t))^2}
e^{ip_{\alpha}(x_1-x_{\alpha})} e^{-\frac {i p_{\alpha}^2 t}{2m}}.
\label {e1}
\end{equation}
Thus, the wave function has two time-dependent parameters,
$x_{\alpha}$ and $p_{\alpha}$.  The total $\emph{n}$-body wave
function is assumed to be a direct product of coherent states:
\begin{equation}
\phi=\phi_{\alpha}
(x_1,x_{\alpha},p_{\alpha},t)\phi_{\beta}(x_2,x_{\beta},
p_{\beta},t)....,         \label {e2}
\end{equation}
where antisymmetrization is neglected. One should, however, keep
in the mind that the Pauli principle, which is very important at
low incident energies, was taken into account. The initial values
of the parameters are chosen in a way that the ensemble
($A_T$+$A_P$) nucleons give a proper density distribution as well
as a proper momentum distribution of the projectile and target
nuclei. The time evolution of the system is calculated using the
generalized variational principle. We start out from the action
\begin{equation}
S=\int_{t_1}^{t_2} {\cal {L}} [\phi,\phi^{*}] d\tau, \label {e3}
\end{equation}
with the Lagrange functional
\begin{equation}
{\cal {L}} =\left(\phi\left|i\hbar \frac
{d}{dt}-H\right|\phi\right), \label {e4}
\end{equation}
where the total time derivative includes the derivatives with
respect to the parameters. The time evolution is obtained by the
requirement that the action is stationary under the allowed
variation of the wave function
\begin{equation}
\delta S=\delta \int_{t_1}^{t_2} {\cal {L}} [\phi ,\phi^{*}] dt=0.
\label{e5}
\end{equation}
If the true solution of the Schr\"odinger equation is contained in
the restricted set of wave function
$\phi_{\alpha}\left({x_{1},x_{\alpha},p_{\alpha}}\right),$ this
variation of the action gives the exact solution of the
Schr\"odinger equation. If the parameter space is too restricted,
we obtain that wave function in the restricted parameter space,
which comes close to the solution of the Schr\"odinger equation.
Performing the variation with the test wave function (2), we
obtain for each parameter $\lambda$ an Euler-Lagrange equation:
\begin{equation}
\frac{d}{dt} \frac{\partial {\cal {L}}}{\partial {\dot
{\lambda}}}-\frac{\partial \cal {L}} {\partial \lambda}=0.
\label{e6}
\end{equation}
For each coherent state and a Hamiltonian of the form, \\

$H=\sum_{\alpha}
\left[T_{\alpha}+{\frac{1}{2}}\sum_{\alpha\beta}V_{\alpha\beta}\right]$,
the Lagrangian and the Euler-Lagrange function can be easily
calculated \cite{aich91}:
\begin{equation}
{\cal {L}} = \sum_{\alpha}{\dot {\bf x}_{\alpha}} {\bf
p}_{\alpha}-\sum_{\beta} \langle{V_{\alpha
\beta}}\rangle-\frac{3}{2Lm}, \label{e7}
\end{equation}
\begin{equation}
{\dot {\bf x}_{\alpha}}=\frac{{\bf
p}_\alpha}{m}+\nabla_{p_{\alpha}}\sum_{\beta} \langle{V_{\alpha
\beta}}\rangle, \label {e8}
\end{equation}
\begin{equation}
{\dot {\bf p}_{\alpha}}=-\nabla_{{\bf x}_{\alpha}}\sum_{\beta}
\langle{V_{\alpha \beta}}\rangle. \label {e9}
\end{equation}
Thus, the variational approach has reduced the $\emph{n}$-body
Schr\"odinger equation to a set of $\emph{6n}$ different equations
for the parameters that can be solved numerically. If one inspects
the formalism carefully, one finds that the interaction potential,
which is actually the Br\"{u}ckner $\emph{G}$-matrix, can be
divided into two parts: (i) a real part and (ii) an imaginary
part. The real part of the potential acts like a potential,
whereas the imaginary part is proportional to the cross section.

In the present model, the interaction potential comprises of the
following terms:
\begin{equation}
V_{\alpha\beta} = V_{loc}^{2} + V_{loc}^{3} + V_{Coul} + V_{Yuk} +
V_{MDI}, \label {e10}
\end {equation}
where $V_{loc}$ is the Skyrme force and $V_{Coul}$, $V_{Yuk}$ and
$V_{MDI}$ define, respectively, the Coulomb, Yukawa and momentum
dependent potentials. The Yukawa term separates the surface which
also plays a role in low-energy processes like fusion and cluster
radioactivity \cite{dutt10,malik}. The expectation value of these
potentials is calculated as
\begin{eqnarray}
V^2_{loc}& =& \int f_{\alpha} ({\bf p}_{\alpha}, {\bf r}_{\alpha},
t) f_{\beta}({\bf p}_{\beta}, {\bf r}_{\beta}, t)V_I ^{(2)}({\bf
r}_{\alpha}, {\bf r}_{\beta})
\nonumber\\
&  & \times {d^{3} {\bf r}_{\alpha} d^{3} {\bf r}_{\beta}
d^{3}{\bf p}_{\alpha}  d^{3}{\bf p}_{\beta},}
\end{eqnarray}
\begin{eqnarray}
V^3_{loc}& =& \int  f_{\alpha} ({\bf p}_{\alpha}, {\bf
r}_{\alpha}, t) f_{\beta}({\bf p}_{\beta}, {\bf r}_{\beta},t)
f_{\gamma} ({\bf p}_{\gamma}, {\bf r}_{\gamma}, t)
\nonumber\\
&  & \times  V_I^{(3)} ({\bf r}_{\alpha},{\bf r}_{\beta},{\bf
r}_{\gamma}) d^{3} {\bf r}_{\alpha} d^{3} {\bf r}_{\beta} d^{3}
{\bf r}_{\gamma}
\nonumber\\
&  & \times d^{3} {\bf p}_{\alpha}d^{3} {\bf p}_{\beta} d^{3} {\bf
p}_{\gamma}.
\end{eqnarray}
where $f_{\alpha}({\bf p}_{\alpha}, {\bf r}_{\alpha}, t)$ is the
Wigner density which corresponds to the wave functions [Eq. 2]. If
we deal with the local Skyrme force only, we get
\small{\begin{equation} V^{Skyrme} = \sum_{{\alpha}=1}^{A_T+A_P}
\left[\frac {A}{2} \sum_{{\beta}=1} \left(\frac
{\tilde{\rho}_{\alpha \beta}}{\rho_0}\right) + \frac
{B}{C+1}\sum_{{\beta}\ne {\alpha}} \left(\frac {\tilde
{\rho}_{\alpha \beta}} {\rho_0}\right)^C\right].
\end{equation}}

\normalsize Here $\emph{A}$, $\emph{B}$ and $\emph{C}$ are the
Skyrme parameters which are defined according to the ground- state
properties of a nucleus. Different values of $\emph{C}$ lead to
different equations of state. A larger value of $\emph{C}$ (= 380
MeV) is often dubbed a stiff equation of state. The finite range
Yukawa ($V_{Yuk}$) potential and effective Coulomb potential
($V_{Coul}$) read as
\begin{equation}
V_{Yuk} = \sum_{j, i\neq j} t_{3}
\frac{exp\{-|\textbf{r}_{\textbf{i}}-\textbf{r}_{\textbf{j}}|\}/\mu}{|\textbf{r}_{\textbf{i}}-\textbf{r}_{\textbf{j}}|/\mu},
\end{equation}
\begin{equation}
V_{Coul} = \sum_{j, i\neq
j}\frac{Z_{eff}^{2}e^{2}}{|\textbf{r}_{\textbf{i}}-\textbf{r}_{\textbf{j}}|}.
\end{equation}
\par
The Yukawa interaction (with $t_{3}$= -6.66 MeV and $\mu$ = 1.5
fm) is essential for the surface effects. The momentum-dependent
interactions (MDI) are obtained by parameterizing the momentum
dependence of the real part of the optical potential. The final
form of the potential reads as follows \cite{aich91}:
\begin{equation}
U^{MDI}\approx
t_{4}\ln^{2}[t_{5}(\textbf{p}_{1}-\textbf{p}_{2})^{2}+1]\delta(\textbf{r}_{1}-\textbf{r}_{2}).
\end{equation}
where $t_{4}$=1.57 MeV and $t_{5}$=5$\times$10$^{-4}$ MeV$^{-2}$.
A parameterized form of the local plus MDI potential is given by
\begin{equation}
U=\alpha(\frac{\rho}{\rho_{o}})+\beta(\frac{\rho}{\rho_{o}})+\delta
\ln^{2}[\epsilon(\rho/\rho_{o})^{2/3}+1]\rho/\rho_{o}.
\end{equation}
The parameters $\alpha$, $\beta$, $\gamma$, $\delta$ and
$\epsilon$ are listed in Ref. \cite{aich91}. The
momentum-dependent part of the interaction acts strongly in the
cases where the system is mildly excited
\cite{kumarpuri99,kumarpuri98}. In this case, the MDI is reported
to generate a lot more fragments compared to the static equation
of state. For a detailed discussion of the different equations of
state and MDI, the reader is referred to Refs.
\cite{blaich93,kumarpuri99,kumarpuri98}. The relativistic effect
does not play a role in the low incident energy of present
interest.
\par
The phase space of the nucleons is stored at several time steps.
The QMD model does not give any information about the fragments
observed at the final stage of the reaction. To construct
fragments from the present phase space, one needs the
clusterization algorithms. We concentrate here on the minimum
spanning tree (MST) method and the minimum spanning tree method
with binding energy check (MSTB) only.
\par
\subsection{Different clusterization methods}
\subsubsection{Minimum spanning tree method}
 The widely used clusterization algorithm is the MST method \cite{jsingh}. In the MST method, two
nucleons are allowed to share the same fragment if their centroids
are closer than a distance $\emph{r}_{min}$,
\begin{equation}
|\textbf{r}_{\textbf{i}}-\textbf{r}_{\textbf{j}}| \leq
\emph{r}_{min},
\end{equation}
where $\textbf{r}_{\textbf{i}}$ and $\textbf{r}_{\textbf{j}}$ are
the spatial positions of both nucleons. The value of
$\emph{r}_{min}$ can vary between 2 and 4 fm. This method cannot
address the question of time scale. This method gives a big
fragment at high density, which splits into several light and
medium mass fragments after several hundred fm/c. This procedure
gives same fragment pattern for times later than 200 fm/c, but
cannot be used for earlier times.
\begin{figure}[!t]
\centering
 \vskip -0.9cm
\includegraphics[angle=0,width=8.9cm, height=13cm]{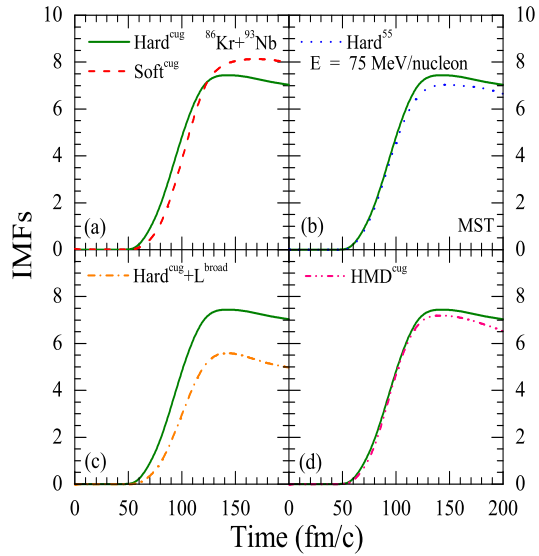}
 \vskip -4.2cm \caption{ (Color online) The time evolution of IMFs (5$\leq$A $\leq$44) for the reaction of $^{86}$Kr+$^{93}$Nb at incident energy of 75
MeV/nucleon for different model ingredients .}\label{fig1}
\end{figure}
\par
\subsubsection{Minimum spanning tree method with binding energy check }
This is an improved version of the normal MST method. First, the
simulated phase space is analyzed with the MST method and
pre-clusters are sorted out. Each of the pre-clusters is then
subjected to a binding energy check \cite{puri09,jsingh}:
\begin{equation}
\zeta_{i} =
\frac{1}{N^{f}}\sum_{i=1}^{N^{f}}[\frac{(\textbf{p}_{i}-P_{N^{f}}^{c.m.})^{2}}{2m_{i}}
+ \frac{1}{2}\sum_{j \neq
i}^{N^{f}}V_{ij}(\textbf{r}_{i},\textbf{r}_{j})]< E_{bind}.
\end{equation}
We take $E_{bind}$ = -4.0 MeV if $N^{f}\geq 3$ and $E_{bind}$ =
0.0 otherwise. Here $N^{f}$ is the number of nucleons in a
fragment and $P_{N^{f}}^{cm}$ is the center-of-mass momentum of
the fragment. This is known as the minimum spanning tree method
with Binding energy check (MSTB) \cite{puri09,jsingh}. The
fragments formed with the MSTB are reliable and stable at early
stages of the reactions.
\par
\section{Results and discussion}
We simulated the central reactions of $^{20}$Ne+$^{20}$Ne
(E$_{lab}$ = 10-55 MeV/nucleon), $^{40}$Ar+$^{45}$Sc (E$_{lab}$ =
35-115 MeV/nucleon), $^{58}$Ni+$^{58}$Ni (E$_{lab}$ = 35-105
MeV/nucleon), $^{86}$Kr+$^{93}$Nb (E$_{lab}$ = 35-95 MeV/nucleon),
$^{129}$Xe+$^{118}$Sn (E$_{lab}$= 45-140 MeV/nucleon),
$^{86}$Kr+$^{197}$Au (E$_{lab}$= 35-400 MeV/nucleon) and
$^{197}$Au+$^{197}$Au (E$_{lab}$ = 70-130 MeV/nucleon). The
energies are guided by experiments \cite{tsang93,sisan01,peas94}.
For the present study, we use hard (labeled Hard), soft (Soft),
hard with MDI (HMD), and soft with MDI (SMD) equations of state.
We also use the standard energy-dependent Cugnon cross section
($\sigma$$_{nn} ^{free}$) \cite{kumarpuri98} and constant
isotropic cross section of 55 mb strength in addition to two
different widths of Gaussian L = 1.08 and 2.16 fm$^{2}$
(L$^{broad}$). The superscripts represent cross section. The phase
space is clusterized using the clusterization methods described
previously. The reactions are followed until 200 fm/c but the
conclusions do not change when the reaction is complete, employing
the validity of both algorithms.
\begin{figure}[!t]
\centering \vskip 0.1cm
\includegraphics[angle=0,width=7cm, height=9.6cm]{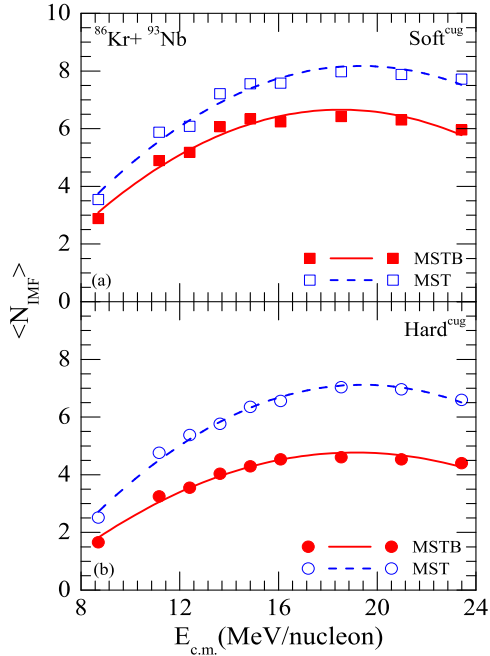}
\vskip -0cm \caption{(Color online) The mean IMF multiplicity,
$\langle$N$_{IMF}\rangle$, as a function of incident energy in
center-of-mass frame, E$_{c.m.}$, for the reaction
$^{86}$Kr+$^{93}$Nb. Solid (dashed) curves show the quadratic fits
to the model calculations for MSTB (MST) to estimate the peak
center-of-mass energy at which the maximal IMF emission
occurs.}\label{fig2}
\end{figure}
\begin{figure}[!t]
\centering \vskip 0cm
\includegraphics[angle=0,width=8cm]{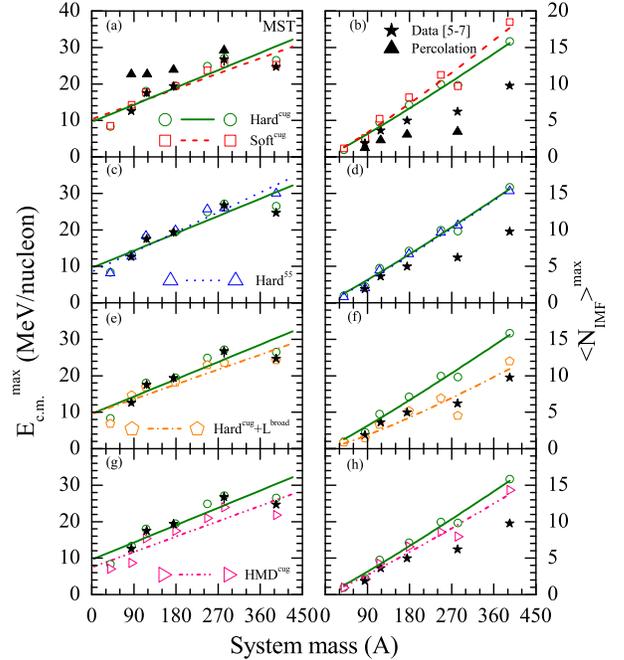}
\vskip -0.6cm \caption{(Color online) The E$_{c.m.} ^{max}$ (left)
and $\langle$N$_{IMF}\rangle^{max}$ (right) as a function of
composite mass of the system (A). The different lines in left
(right) panels represent the linear fits (power-law fits).
Comparison of model calculations is made with experimental data
\cite{tsang93,sisan01,peas94} (solid stars). The percolation
calculations \cite{sisan01} (solid triangles) are also shown in
figure.}\label{fig3}
\end{figure}
\par
In fig. 1, we display the time evolution of IMFs for the reaction
$^{86}$Kr+$^{93}$Nb at incident energy of 75 MeV/nucleon employing
the MST method. In fig. 1(a), we display the model calculations
using Hard$^{Cug}$ (solid line) and Soft$^{Cug}$ (dashed line).
From fig. 1(a), we see that the number of IMFs is larger in case
of Soft compared to that of Hard. This is because of the fact that
soft matter is easily compressed, resulting in greater achieved
density, which in turn leads to the large number of IMFs compared
to that in the Hard. It is worth mentioning here that the effect
could be opposite at higher energies,because at higher energies
the IMFs may further break into LCPs and free nucleons. In fig.
1(b), we display the results for Hard$^{Cug}$ and Hard$^{55}$
(dotted line). As evident from the fig. 1(b), the number of IMFs
is nearly same for both types of cross sections. This may be
because, for the central collisions, since the excitation energy
is already high,different cross sections have a negligible role to
play. In fig. 1(c), we display the results for the Hard case along
with two different widths of the Gaussian,that is, L and
L$^{broad}$ (dash-dotted line). We find that the width of Gaussian
has a considerable impact on fragmentation. As we change the
Gaussian width (L) from 4.33 to 8.66 fm$^{2}$, the multiplicity of
IMFs is reduced by $\approx$ 30$\%$. Interestingly, the kaon yield
also gets reduced by the same amount \cite{hart98}. Owing to its
large interaction range, an extended wave packet (i.e.
L$^{broad}$) connects a large number of nucleons in a fragment, as
a result, it generates heavier fragments compared to what is
obtained with a smaller width. It is worth mentioning here that
the width of the Gaussian has a considerable effect on the
collective flow \cite{gautm10,hart98} as well as on pion
production \cite{hart94,hart98}. In fig. 1(d), we display the
results using Hard and HMD (dash-dot-dotted line). Again the
number of IMFs are nearly same for both equations of state (EOS).
This is expected because the effect of MDI is small at these
energies. However, the scenario is completely different at high
energies; at high energies, owing to the repulsive nature of MDI,
there is a large destruction of initial correlations and the
additional momentum dependence further destroys the correlations
reducing further the multiplicity of IMFs. This leads to the
emission of lots of nucleons and LCPs \cite{goyal09}.

\par
In fig. 2, we display the average multiplicity of IMFs,
$\langle$N$_{IMF}\rangle$, as a function of incident energy in the
center-of-mass frame (E$_{c.m.}$) for the $^{86}$Kr+$^{93}$Nb
reaction employing the MST (open symbols) and the MSTB (solid
symbols) methods. Figures. 2(a) and 2(b) are for Soft$^{Cug}$ and
Hard$^{Cug}$, respectively. Lines represent the quadratic fit to
the model calculations. In both cases, the number of IMFs first
increases with incident energy, attains a maximum, and then
decreases, in agreement with previous studies
\cite{tsang93,sisan01,peas94,ogil91,puri09}. Clearly,
$\langle$N$_{IMF}\rangle$ is greater for the MST method than for
MSTB, because in the case of the MSTB, along with spatial
correlations, an additional check for binding energy is also used;
therefore, it filters out the loosely bound fragments which decay
later. Hence, the fragments obtained with THE MSTB are properly
bound. A similar trend is obtained for all other reactions as well
as different model ingredients used in the present study but is
less pronounced in lighter systems like $^{20}$Ne+$^{20}$Ne and
$^{40}$Ar+$^{45}$Sc as compared to heavier systems. However, for
the Gaussian width L$^{broad}$, the value of
$\langle$N$_{IMF}\rangle$ is nearly zero in this incident energy
range using THE MSTB (not shown here). This is because an extended
wave packet (i.e. L$^{broad}$) connects a large number of nucleons
in a fragment; as a result, it generates heavier fragments and the
additional binding energy check further excludes the unbound
fragments.
\par
 In fig. 3, we display the peak center-of-mass energy E$_{c.m.}
^{max}$ (left panels) and peak multiplicity of IMFs
$\langle$N$_{IMF}\rangle^{max}$ (right panels) as a function of
the combined mass of the system employing the MST method. In the
left panels, lines represent linear fitting proportional to $mA$
and in the right panels, lines represent power-law fitting
proportional to A$^{\tau}$. The multiplicity of IMFs, in the case
of $^{20}$Ne+$^{20}$Ne and $^{40}$Ar+$^{45}$Sc, is obtained by
excluding the largest and second largest fragment, respectively,
to get the accurate information about the system size dependence.
$\langle$N$_{IMF}\rangle^{max}$ and corresponding E$_{c.m.}
^{max}$ are obtained by making a quadratic fit to the model
calculations for $\langle$N$_{IMF}\rangle$ as a function of
(E$_{c.m.}$). From the left panels, we find that the mass
dependence of E$_{c.m.} ^{max}$ is insensitive to different EOS
(fig. 3a), nn cross section (fig. 3b), as well as the width of the
Gaussian [fig. 3c]. It is slightly sensitive to MDI because, for
heavy systems, the value of E$_{c.m.} ^{max}$ is greater, as a
result of which is that the effect of MDI becomes non-negligible
and creates the $\langle$N$_{IMF}\rangle^{max}$ at smaller
energies. From the right panels where we display the mass
dependence of $\langle$N$_{IMF}\rangle^{max}$, we find that the
peak multiplicity is insensitive to cross section (fig. 3d) and
MDI (fig. 3h) (for explanation see discussion of fig. 1). It is
slightly sensitive to the EOS (fig. 3b) but highly sensitive to
the Gaussian width (fig. 3f). On increasing the width of the
Gaussian, $\langle$N$_{IMF}\rangle^{max}$ reduces to a large
extent. As discussed earlier, an extended wave packet (i.e.
L$^{broad}$) will connect a large number of nucleons in a
fragment; as a result, it generates heavier fragments compared to
what one obtains with smaller width. From fig. 3, we see that
E$_{c.m.} ^{max}$ shows linear dependence ($\propto$ mA) whereas
$\langle$N$_{IMF}\rangle^{max}$ follows power-law behavior
($\propto$ A$^{\tau}$) with $\tau$ nearly equal to unity.
\par
\begin{figure}[!t]
\centering \vskip -0.8cm
\includegraphics[angle=0,width=8.9cm, height=11.5cm]{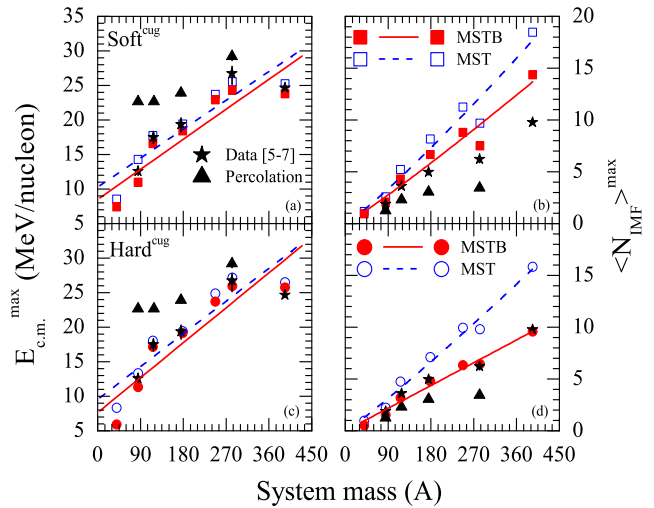}
\vskip -3.4cm \caption{(Color online) The E$_{c.m.} ^{max}$ (left
panels) and $\langle$N$_{IMF}\rangle^{max}$ (right panels) as a
function of composite mass of the system (A) using Soft$^{Cug}$
(upper panels) and Hard$^{Cug}$ (lower panels) employing THE MSTB
and THE MST methods. Lines have same meaning as in fig. 3.
Comparison of model calculations is made with experimental data
\cite{tsang93,sisan01,peas94} (solid stars).}\label{fig4}
\end{figure}

\begin{figure}[!t] \centering
 \vskip 0cm
\includegraphics[angle=0,width=8.5cm]{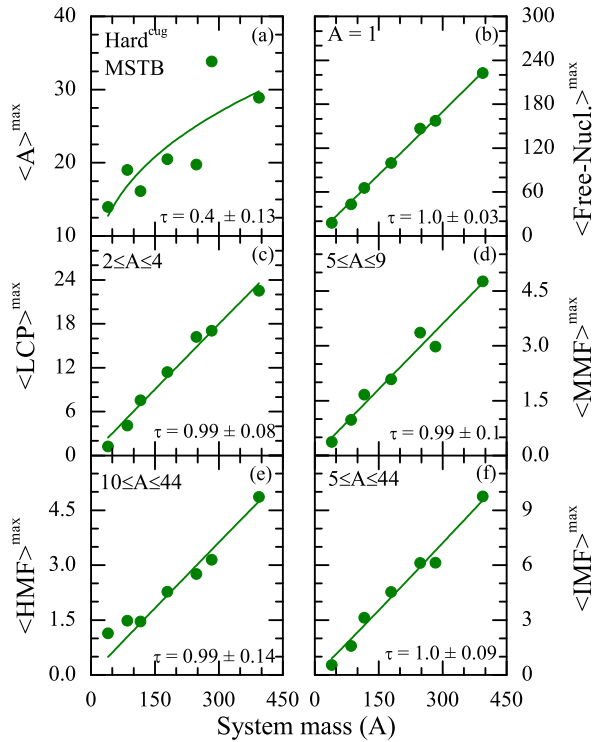}
 \vskip -0.8cm \caption{ (Color online) The largest fragment and multiplicities of free-nucleons, LCPs, MMFs, HMFs, and IMFs as a
 function of composite mass of the colliding nuclei (A) for different reactions at their respective E$_{c.m.}^{max}$ (solid circles).
 Lines represent the power-law fits ($\propto$ A$^{\tau}$).}\label{fig5}
\end{figure}
In fig. 3, the model calculations are also compared with
experimental data \cite{tsang93,sisan01,peas94}. It is clear from
fig. 3 that model calculations for E$_{c.m.} ^{max}$ agree with
experimental data \cite{tsang93,sisan01,peas94} whereas in the
case of $\langle$N$_{IMF}\rangle^{max}$, as the system mass
increases, the difference between model calculations and
experimental results continues to  increase. This behavior is
consistent for all the different choices of model ingredients.
This may be because the fragments obtained with the MST method are
not reliable because this method makes sense only when matter is
diluted and well separated. This is true only in the case of high
beam energy and in central collisions. Therefore, we have to look
for other methods of clusterization. As mentioned earlier, the
fragments obtained with the MSTB are properly bound and reliable.
So, as a next step, we check the system size dependence of
E$_{c.m.} ^{max}$ and $\langle$N$_{IMF}\rangle^{max}$ by using the
MSTB for clusterization.

\par
In fig. 4, we display the E$_{c.m.} ^{max}$ (left panels) and
$\langle$N$_{IMF}\rangle^{max}$ (right panels) for Soft$^{Cug}$
(upper panels) and Hard$^{Cug}$ (bottom panels) as a function of
the combined mass of the system. Solid (open) symbols represent
the MSTB (MST method). From the left-hand panels, we find that
E$_{c.m.} ^{max}$ remains insensitive to the choice of
clusterization method. The same is true for
$\langle$N$_{IMF}\rangle^{max}$ (right-hand panels) but in
low-mass region. As the system mass increases,
$\langle$N$_{IMF}\rangle^{max}$ becomes more and more sensitive to
the method of clusterizaton. The MSTB excludes the loosely bound
fragments, thus reducing the peak IMF multiplicity. The effect is
uniform both for the EOS and for different cross sections (not
shown here).

\par
In fig. 5, we display peak multiplicity (obtained by employing the
MSTB) as a function of composite mass of the system for various
fragments consisting of the largest fragment (A$^{max}$) (fig.
5a), free-nucleons (1$\leq$A $\leq$1) (fig. 5b), light charged
particles (LCPs) (2$\leq$A $\leq$4) (fig. 5c), medium mass
fragments (MMFs) (5$\leq$A $\leq$9) (fig. 5d), heavy mass
fragments (HMFs) (10$\leq$A $\leq$44) (fig. 5e) and intermediate
mass fragments (IMFs) (5$\leq$A $\leq$44) (fig. 5f) for
Hard$^{Cug}$. Lines represent the power-law fitting proportional
to A$^{\tau}$.
\begin{figure}[!t]
\centering
 \vskip 0cm
\includegraphics[angle=0,width=8cm, height=10.5cm]{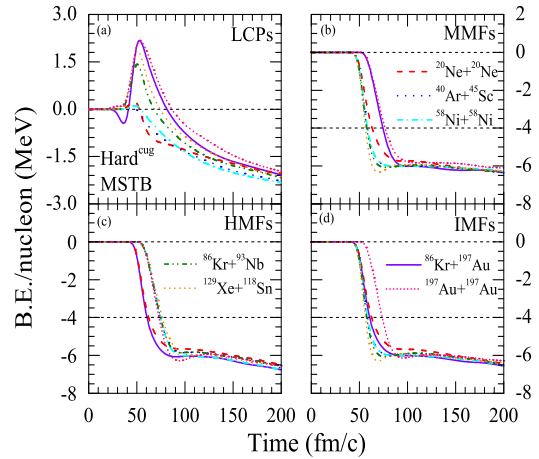}
 \vskip -3cm \caption{ (Color online) The time evolution of binding energy per nucleon for LCPs, MMFs, HMFs, and IMFs.
 Here reactions of $^{20}$Ne+$^{20}$Ne, $^{40}$Ar+$^{45}$Sc, $^{58}$Ni+$^{58}$Ni, $^{86}$Kr+$^{93}$Nb, $^{129}$Xe+$^{118}$Sn, $^{86}$Kr+$^{197}$Au, and $^{197}$Au+$^{197}$Au
 are simulated for central geometry at their corresponding E$_{c.m.}^{max}$.}\label{fig6}
\end{figure}
Interestingly, the peak multiplicities of different fragments
follow a power-law ($\propto$ A$^{\tau}$). The power-law factor
$\tau$ is almost unity in all cases except A$^{max}$ for which
there is no clear system size dependence. The system size
dependence of various fragments was also predicted by Dhawan and
Puri \cite{dhawan06}. Their calculations at the energy of
vanishing flow (i.e., the energy at which the transverse flow
vanishes) clearly suggested the existence of a power-law system
mass dependence for various fragment multiplicities.

\par
To check the stability of fragments, we display in fig. 6, the
binding energy per nucleon as a function of time for LCPs, MMFs,
HMFs, and IMFs. The reactions of $^{20}$Ne+$^{20}$Ne,
$^{40}$Ar+$^{45}$Sc, $^{58}$Ni+$^{58}$Ni, $^{86}$Kr+$^{93}$Nb,
$^{129}$Xe+$^{118}$Sn, $^{86}$Kr+$^{197}$Au and
$^{197}$Au+$^{197}$Au are simulated at laboratory energies
corresponding to their E$_{c.m.} ^{max}$ values, which are
approximately 24, 46, 69, 78, 96, 124, and 105 MeV/nucleon,
respectively. We find that, even at 200 fm/c, small fragments are
still not cold and take a very long time to cool down, whereas the
heavy fragments are properly bound, having binding energy per
nucleon around -5 to -7 MeV.
\par
\section{Summary}
We simulated the central reactions of nearly symmetric, and
asymmetric systems over the entire periodic table at different
incident energies for different EOS, nn cross sections, and
different widths of Gaussians. We observed that the multiplicity
of IMFs ($3\leq$Z$\leq20$) shows a rise and fall with increase in
beam energy in the center-of-mass frame, as already predicted
experimentally and theoretically. We also studied the system size
dependence of peak center-of-mass energy E$_{c.m.} ^{max}$ and
peak IMF multiplicity $\langle$N$_{IMF}\rangle^{max}$. It was
observed that E$_{c.m.}^{max}$ increases linearly with system mass
whereas a power-law ($\propto$ A$^{\tau}$) dependence was observed
for $\langle$N$_{IMF}\rangle^{max}$ with $\tau\sim$1.0. We
compared the system size dependence of E$_{c.m.} ^{max}$ and
$\langle$N$_{IMF}\rangle^{max}$
 for MST and MSTB methods and found that MSTB reduces the $\langle$N$_{IMF}\rangle^{max}$ especially in heavy systems because loosely bound fragments get excluded in MSTB.
 The power-law dependence is also observed for fragments of
different sizes at the energy for which the production of IMFs is
a maximum and the power-law parameter $\tau$ is found to be close
to unity in all cases except A$^{max}$. The stability of fragments
is also checked through binding energy per nucleon. We observed
that, at 200 fm/c, small fragments are still not cold and they
take a very long time to cool down, whereas the heavy fragments
are properly bound.
\section{Acknowledgements}
The work is supported by Indo-French project no. 4104-1.

\end{document}